\documentstyle[psfig,twocolumn,aps]{revtex}

\begin{document}

\draft
\title{Self-Trapped Exciton Defects in a Charge Density Wave:
Electronic Excitations of BaBiO$_3$}

\author{ Philip B. Allen and Ilka B. Bischofs\cite{byline1}}
\address{Department of Physics and Astronomy, State University of New York,
Stony Brook, New York 11794-3800}

\date{\today}

\maketitle

\begin{abstract}

In the previous paper, it was shown that holes doped into
BaBiO$_3$ self-trap as small polarons and bipolarons.
These point defects are energetically favorable partly
because they undo locally the strain in the 
charge-density-wave (Peierls insulator) ground state.
In this paper the neutral excitations of the same model are
discussed.  The lowest electronic excitation is predicted to
be a self-trapped exciton, consisting of an electron and a
hole located on adjacent Bi atoms.  This excitation has
been seen experimentally (but not identified as such) via
the Urbach tail in optical absorption, and the multi-phonon
spectrum of the ``breathing mode'' seen in Raman scattering. 
These two phenomena occur because of the Franck-Condon effect
associated with oxygen displacement in the excited state.

\end{abstract}


\section{Introduction}

The previous paper on BaBiO$_3$ \cite{Bischofs} presents
a simple and yet sufficiently realistic model \cite{Rice} which gives
a microscopic description of the Peierls distortion,
and makes unambiguous predictions about the low-energy
behavior.  The previous paper looks carefully at the
small polarons created by doping or
photoemission.  These were noticed already by Yu, Chen,
and Su \cite{Yu}.  Here we examine what happens
when an electron-hole pair is inserted, and discover a
self-trapped exciton which has not previously been discussed.
This excitation gives a natural explanation for the Urbach tail
seen in optical absorption \cite{Kozlov} and for the multi-phonon 
replicas of the oxygen breathing phonon (the Peierls mode)
seen by Kozlov {\sl et al.} \cite{Kozlov} in infrared absorption, and
by Tajima {\sl et al.} \cite{Tajima} in Raman scattering.

The model \cite{Bischofs}, 
based on work by Rice and Sneddon \cite{Rice}, has nearest-neighbor
hopping (with strength $t$) of electrons between 
Bi $s$-orbitals on a cubic lattice.
Oxygens which lie between pairs of Bi atoms, experience a linear
repulsive force $g(n_2 -n_1)$ away from the Bi atom with the
larger electron charge ($n_1$) and toward the Bi atom with the
smaller electron charge ($n_2$).  The oxygen atom also has a 
harmonic Einstein spring constant $K$.  The model has a single
dimensionless constant $\Gamma=g^2/Kt$ in adiabatic approximation
(motion of the oxygen atoms neglected.)  We choose parameters
$t$=0.35 eV, $K=19$ eV$/\AA^2$, $g=1.39$ eV$/\AA$, yielding
$\Gamma=0.30$.

\section{Atomic Limit $\protect{t}=0$}

As in the previous paper, the zero bandwidth ($t=0$, $\Gamma \rightarrow
\infty$) case provides
a soluble limit which gives good insight into the more complete theory.
If a hole is created in the lower Peierls or occupied valence band,
and an electron in the upper Peierls or empty conduction band,
without allowing any lattice relaxation, the energy cost is
$2\Delta=24g^2/K$.  The hole is a Bi$^{4+}$ ion on a site
normally occupied by Bi$^{3+}$, and the electron is a Bi$^{4+}$
ion on a site normally occupied by Bi$^{5+}$.  

Because of the simple Hamiltonian, the problem has particle-hole symmetry.
This is true at all values of parameters, not just $t=0$.
Upper and lower Peierls bands have energies $\pm \lambda(\vec{k})$
where $\lambda(\vec{k})=\sqrt{\Delta^2+\epsilon(\vec{k})^2}$, and
$\epsilon(\vec{k})$ is the nearest neighbor ``cosine'' band dispersion.
Thus particle and hole bands are symmetrical.

If the electron and hole are spatially separated, each will lower its energy by
lattice relaxation.  Both electron and hole save $3\Delta/4$ in
elastic energy by moving surrounding oxygens half way back to their
cubic perovskite positions.  However, each unpaired electron on a
Bi$^{4+}$ ion has its on-site energy increased by $\Delta/2$, for a net
lowering of energy by $\Delta/4$ for each, or a cost of 3$\Delta/2$
to create the isolated pair of polarons.  

\begin{figure}[t]
\centerline{\psfig{figure=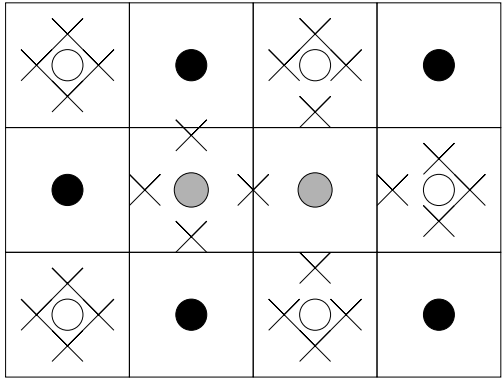,height=2.16in,width=2.88in,angle=0}}
\caption{Schematic structure of the self-trapped exciton. 
The electron is in the left-central cell (``exciton-electron''), 
the hole (``exciton-hole'') in the
right-central cell.  Filled circles are 
Bi$^{3+}$ ions with two $s$ electrons, shaded circles are
Bi$^{4+}$ ions with one $s$ electron, and open circles are
Bi$^{5+}$ ions with no $s$ electrons.  X's denote oxygen ions.
The central oxygen ion is undisplaced, which lowers the energy relative
to separated electron and hole polarons.}
\label{fig:str}
\end{figure}

As shown in Fig. \ref{fig:str}, there is an advantage to having
the electron and hole polarons form on adjacent Bi ions.  The
oxygen between the two Bi$^{4+}$ ions can further relax to the
undistorted perovskite position.  Rather than 12, now only 10 oxygens
are half displaced, saving an additional $\Delta/12=g^2/K$ in elastic energy.
This bound pair of polarons, called a ``self-trapped exciton,''
costs $E_{\rm exc}=17\Delta/12$ rather than $2\Delta$ to create.  
The total energy gain 
from lattice relaxation is the exciton binding energy 
$\epsilon_{\rm exc}=7g^2 /K$.
Taking the on-site Coulomb interaction into
account, the break-up of the Bi$^{3+}$ paired electrons
into separated electron-hole pairs will reduce the
excited state energy further, to $\epsilon_{\rm exc}=7g^2 /K + U$ where  
$U$ is the Hubbard on-site Coulomb repulsion.

\par
\par

\begin{figure}[t]
\centerline{\psfig{figure=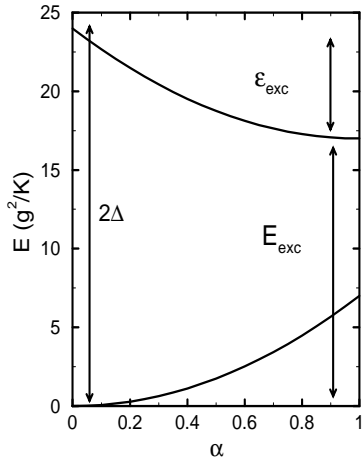,height=2.6in,width=2.0in,angle=0}}
\caption{Excited state and ground state energy versus configuration 
coordinate $\alpha$ for $t=0$. At
$\alpha=0$, the ground and excited state are separated by the gap $2\Delta$. 
The self-trapped exciton at $\alpha=1$ has its energy lowered by the 
maximal amount -- the exciton binding energy $\epsilon_{\rm exc}$. 
Including on-site Coulomb repulsion, the energy curve of the excited 
state will be shifted down by the Hubbard $U$.}
\label{fig:pot}
\end{figure}

The energy of the ground and excited state depend on the amount of lattice 
distortion. The configuration
coordinate $\alpha$ smoothly connects the optimal oxygen configurations of 
the ground state at $\alpha=0$
and the self-trapped exciton state at $\alpha=1$, as shown in 
Fig. \ref{fig:pot}.  The energy depends 
quadratically on $\alpha$, separating ground and excited state by the 
gap $2\Delta$ at $\alpha=0$.  

\par

\section{Numerical Results}

For $t\neq 0$ we are not able to solve analytically, and rely on
numerical studies.  We find that
the excited state has zero slope $dE/d\alpha=0$ at $\alpha=0$, and
downward curvature.  
No potential barrier separates the
excited state at $\alpha=0$ and the self-trapped exciton at 
$\alpha=1$.  This is contrary to most 
self-trapped exciton theories \cite{Song}, 
where a metastable free excited state and a self-trapped 
exciton state can coexist, separated by a potential barrier.

For $t\neq 0$, localizing and delocalizing terms in the Hamiltonian compete 
against each other, yielding a critical coupling constant $\Gamma_c(\rm exc)$ 
above which self-trapped exciton states exist. To obtain the lowest excited 
state numerically, oxygen positions are varied using
the algorithm described in the preceding paper, under the 
constraint that an electron was removed from the highest-lying valence 
band state and inserted in the lowest-lying conduction band state.
We find $\Gamma_c(\rm exc)\approx 0.175$, close to the polaron 
stability limit $\Gamma_c(P)=0.18$ of the previous paper. 
Since the self-trapped exciton is an electron-polaron weakly 
bound to a hole-polaron,
it is not surprising that the polaron and the self-trapped exciton 
have similar threshold coupling constants.  
The first correction to strong coupling for $\epsilon_{\rm exc}$ 
can be approximated using 
a ``vacuum correction'' of order $1/\Gamma^2$ discussed in Sect.III.A of 
the preceding paper: $\epsilon_{\rm exc}/t \approx 7\Gamma 
-1/2 \Gamma$ and approaches asymptotically the $t=0$ limit 
$\epsilon_{\rm exc}=7g^2 /K$.
\par
\begin{figure}[t]
\centerline{\psfig{figure=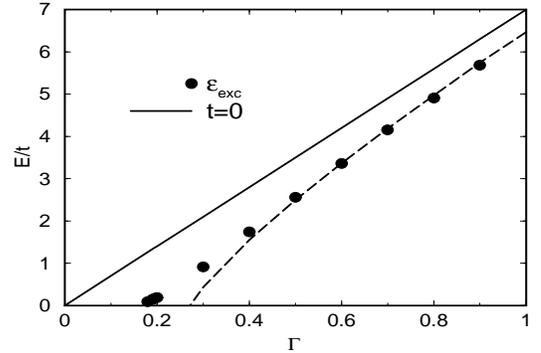,height=1.9in,width=2.88in,angle=0}}
\caption{Exciton binding energy $\epsilon_{\rm exc}$ for $t\neq 0$ 
computed numerically. The bold solid line is the $t=0$ prediction. 
The dashed line is an approximate $1/\Gamma^2$ correction.}
\label{fig:be}
\end{figure}
\begin{figure}[t]
\centerline{\psfig{figure=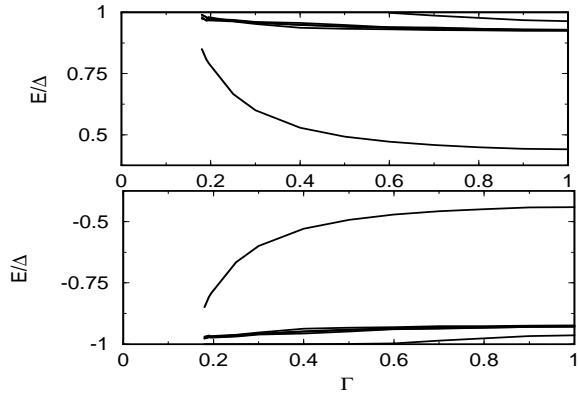,height=2.1in,width=3.0in,angle=0}}
\caption{Energies of localized single particle states which accompany
the self-trapped exciton states
lying in the gap. The spectrum has particle-hole
symmetry. Five doubly occupied states split from the valence band.
corresponding to states surrounding the exciton-electron. 
The mid-gap 
states derived from the valence and conduction band respectively are 
occupied singly each, corresponding to the exciton-hole and 
exciton-electron.  Five empty states split from the conduction
band.  The ten states near the conduction and valence band
edges are localized on the 10 Bi atoms which
surround the two Bi atoms which constitute the exciton.} 
\label{fig:gapst}
\end{figure}

The behavior of self-trapped exciton states and
their evolution with $\Gamma$ is very 
similar that of polarons \cite{Bischofs}.
The local change in the oxygen environment leads to the appearance of 
localized gap states (Fig.\ref{fig:gapst}) in addition to the
states which constitute the exciton.
\par
\begin{figure}[t]
\centerline{\psfig{figure=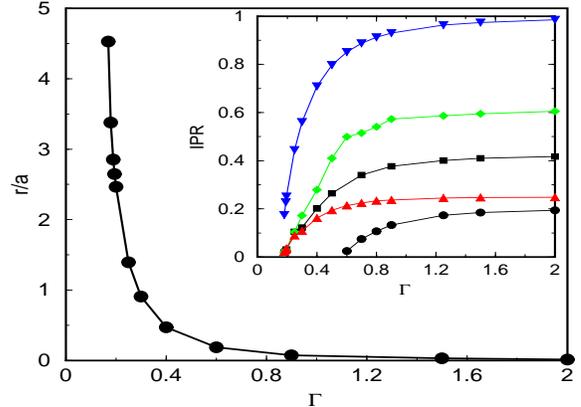,height=2.2in,width=3.0in,angle=0}}
\caption{Radius  $\langle r \rangle$  of the exciton-hole and 
exciton-electron.  A radius of $0$ corresponds to complete 
localization of the wave-function at one site. The radius diverges 
as $\Gamma\rightarrow\Gamma_c(E)$. The finite value at 
$\Gamma_c(E)$ is caused by the finite system size of the numerical 
calculations.  Inset: IPR-values of the gap states, indicating 
their degree of localization. They approach their asymptotic (symmetry 
determined) IPR values of $\frac{1}{5}$, $\frac{1}{4}$, 
$\frac{1}{2}$ (doubly degenerate), $\frac{13}{20}$ and $1$ for 
$\Gamma\rightarrow\infty$. }
\label{fig:rad}
\end{figure}

Six states emerge from the valence band and six from the conduction band, 
symmetrically around $E=0$, because of electron-hole symmetry.
The lowest five states
have a small upward shift ($g^2/K$ when $t=0$) 
from the top of the valence band, and remain doubly occupied.
These are electrons on the five A sites surrounding the B 
sublattice site now occupied by the excited electron (called the 
``exciton-electron'').  Their degeneracy is partly lifted 
when they couple to other sites by hopping, shifting their energies 
toward the valence band.  The next state up also derives from the
valence band, and is occupied once.  It is the electron
that remains behind on an A site when one electron is removed.
Alternately, this is the site of the ``exciton-hole''.  Its energy
($7g^2/K$ above the valence band when $t$=0) is slightly higher
than for an isolated hole state ($6g^2/K$) because one of the six
surrounding oxygens, shared by the exciton-electron and
the exciton-hole, is fully relaxed to the symmetric position.
With decreasing coupling strength, 
the states spread out, as can be seen in the continuous evolution 
of their inverse participation ratios (IPR) in the inset in 
Fig. \ref{fig:rad} . The IPR $1/P_i$ for a state $i$ is defined 
as 
\begin{equation}
\frac{1}{P_i}= \sum_l |\Psi_i(r_l)|^4
\label{eq:IPR}
\end{equation}
where the sum runs over all 
sites.  It measures the localization of a state.  An IPR of $1/P$ 
represents a state that is localized on $P$ atoms.
The mean radius $\langle r \rangle$ of the two states comprising the
exciton is
shown in Fig. \ref{fig:rad}.  The particles are confined approximately to 
one site for $\Gamma>0.4$. 
For $\Gamma<0.4$, they  delocalize rapidly, as indicated by the 
diverging radius.  The self-trapped exciton
evolves smoothly into a localized amplitude
excitation of the charge-density wave, before
disappearing into a band electron-hole pair at
$\Gamma_c(E)$, just as \cite{Bischofs} the small polaron
evolves through a large CDW-like polaron state before becoming
a band hole at $\Gamma_c(P)$.
\par

\section{Optical Properties}

In this section we discuss the observability of self-trapped excitons
in experiment {\it via}
an Urbach tail in optical absorption experiments and a 
multi-phonon spectrum of the ``breathing mode'' in phonon 
experiments, both governed by the Franck-Condon effect.
Zheng and Nasu \cite{Zheng} have also obtained theoretically a tail
in the optical conductivity.  Their theory includes quantum lattice
fluctuations (as does ours) and bears some relation to our theory.

\begin{figure}[t]
\centerline{\psfig{figure=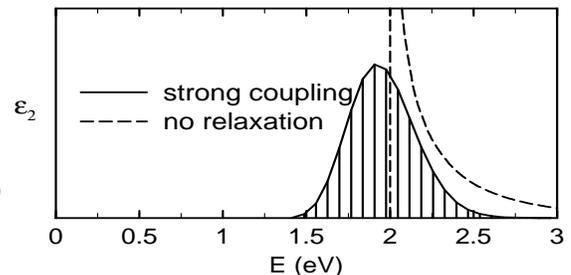,height=1.5in,width=3.0in,angle=0}}
\caption{Imaginary part of dielectric constant versus frequency. 
The dashed line shows Eq.(\protect\ref{eq:nr}), 
where the integral was evaluated with a 
tetrahedron code \protect\cite{Leh,Jep}. Taking lattice relaxation into 
account the absorption spectra will be dominated by the Franck-Condon 
effect Eq.(\protect\ref{eq:FC}), 
leading to significant peak broadening and absorption 
in the gap, known as an Urbach tail.}
\label{fig:e2lin}
\end{figure}

As a first step we neglect 
both Coulomb and lattice polarization effects.  Then the imaginary 
part of the dielectric constant can be computed exactly in electric 
dipole approximation.  The dipole matrix element between
an $s$ state on a Bi atom at $\vec{R}$ and one on the neighboring
Bi atom at $\vec{R}+a{\hat x}$ is $-imta{\hat x}/\hbar.$
This gives the correct intraband $<\vec{k}|\vec{p}|\vec{k}>
=(m/\hbar)\partial \epsilon/\partial \vec{k}$ relation between
momentum matrix element and group velocity.  Then for
for unpolarized light we get
\begin{eqnarray}
\epsilon_2(\omega)&=&\frac{4 e^2 t^2}{m^2 \omega^2\pi \hbar^2}
\int_{BZ}d^3\vec k(\sin^2(k_xa)+\sin^2(k_ya)+\sin^2(k_z a))\nonumber \\
&\times& \delta(2\sqrt{\epsilon_k^2+\Delta^2}-\hbar \omega).
\label{eq:nr}
\end{eqnarray}
This was evaluated numerically using a tetrahedron code \cite{Leh,Jep},
and is shown as the dashed line in Fig.\ref{fig:e2lin}. 
The behavior of $\epsilon_2$ is dominated by the 
divergence of the joint-density of states at photon energy 
$\hbar\omega=2\Delta$,
corresponding to transitions from states at the top of the 
lower Peierls band into those at the bottom of the upper band. 
These states are mainly
responsible for the charge disproportion and have vanishing amplitude on 
the $B$ and $A$ sites respectively.  
These are also the states from which localized polaron and self-trapped
exciton states form.  Since transitions between 
these states are highest in optical
density, we may expect the line shape to be dominated by the 
signature of self-trapped excitons.  

Materials with self-trapped excitons show Franck-Condon broadening 
of electronic transitions, a concept well known to molecular
spectroscopists, since molecular excited electronic states generally 
have altered nuclear coordinates.  The process involves
simultaneous creation
of electronic excitation and nuclear vibrations, causing the
appearance of vibrational sidebands in the 
electronic spectra \cite{Herz}.  In simple solids, excitations are 
usually delocalized which eliminates this effect\cite{Ras}. However, 
in case of self-localized states, because of their significant lattice 
relaxation, Franck-Condon broadening of the electronic transitions 
must reappear. $\epsilon_2(\omega)$ in Franck-Condon approximation is 
given by:
\begin{eqnarray}
\epsilon_2(\omega) \propto |P_{\lambda \lambda^{\prime}}|^2 
\sum_n \sum_{n^{\prime}}
 w_{\lambda n} |\langle \lambda n|\lambda^{\prime} n^{\prime}\rangle |^2 
\delta(\hbar \omega + E_{\lambda n} - E_{\lambda^{\prime} n^{\prime}}),
\label{eq:FC}
\end{eqnarray} 
where $P_{\lambda \lambda^{\prime}}$ is the electronic dipole matrix 
element between the electronic ground $ \lambda$ and excited state 
$ \lambda^{\prime}$, $w_{\lambda n}$ the probability to find $n$ 
vibrational quanta in the electronic ground state and 
$\langle \lambda n|\lambda^{\prime} n^{\prime}\rangle$ the overlap 
integral between the vibrational wave functions $\chi_{\lambda n}$.
This integral can be large even when the vibrational quantum numbers
$n$, $n^{\prime}$ are quite different from each other, because the 
vibrational wave-functions $\chi_{\lambda n}$,
$\chi_{\lambda^{\prime} n^{\prime}}$ have origins which
are off-set from each other.

At zero temperature, the broadening of the electronic transition 
derives from the overlap of zero point
motion of the oxygens with excited vibrations in the electronic
excited state.  The sum over $n$ in Eq.(\ref{eq:FC}) then
has only one term, $w_{\lambda n}=\delta_{n,0}$ .
We lift the adiabatic approximation applied so far by adding 
quantized vibrations, but treat the problem in the atomic 
limit ($t=0$), i.e. in single site approximation. The excited electron 
states couple to eleven oxygens, 
which are treated as independent Einstein oscillators with frequency 
$\omega_0$ and oxygen mass $M$ in our model. They move in harmonic 
potentials and thus the  Franck-Condon factor \cite{Huang} is a product of 
eleven vibrational overlap integrals of the form 
$|\langle 0|n\rangle|^2$ with:
\begin{eqnarray}
\langle 0|n\rangle =\left(\frac{\alpha^2}{2}\right)^{\frac{n}{2}} 
\frac{(R-R^{\prime})^n}{\sqrt{n!}} 
\exp(-\frac{\alpha^2}{4}(R-R^{\prime})^2),
\label{eq:Huang}
\end{eqnarray}
where $\alpha=\sqrt{M\omega_0/\hbar}$ and $R$ and $R^{\prime}$ are
the oxygen equilibrium positions of the
ground and excited state respectively. 
The total absorption at a frequency $\hbar \omega=
E_{\rm exc}+n\hbar \omega_0$,
requires a sum over all different ways to distribute 
$n$ quanta over the eleven oxygens.  The result is
\begin{eqnarray}
\epsilon_2(\omega) &=&\frac{4 \pi^2 e^2}{m^2 \omega^2} \sum_{n}
\frac{1}{n!}\left(\frac{7\Delta}{12\hbar\omega_0}\right)^{n} 
\exp(-\frac{7\Delta}{12\hbar\omega_0}) \nonumber\\
&\times&\delta(\hbar\omega -\frac{17}{12}\Delta +n \hbar \omega_0).
\label{eq:FCT0}
\end{eqnarray}
In our simplified Einstein model, absorption occurs at discrete frequencies 
corresponding to
different numbers of vibrations created in the absorption process. 
In reality these peaks are broadened by phonon dispersion, and 
discreteness of the individual peaks should only be seen
for one or two vibrational quanta.
The line shape should look similar to the 
envelope function 
of Fig.\ref{fig:e2lin}, with maximal absorption at $\omega=2\Delta$ 
and significant absorption below the gap at $2\Delta$=2 eV.

\par
\begin{figure}[t]
\centerline{\psfig{figure=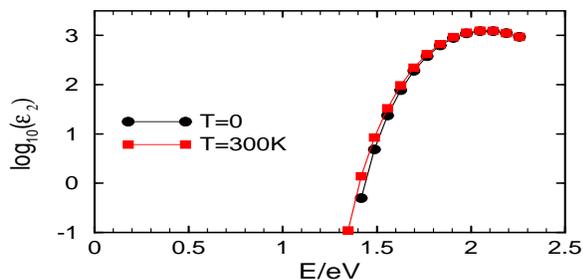,height=1.5in,width=3.0in,angle=0}}
\caption{Imaginary part of dielectric constant (logarithmic scale)
{\it versus} frequency at $T$=0, and
the first temperature correction to the 
Franck-Condon spectra at 300K.}
\label{fig:e2log}
\end{figure}

The absorption spectra of self-trapped excitons will change with temperature 
due to the
temperature-dependent occupation of vibrational states in the 
electronic ground state given by $w_{\lambda,n}= 
\exp(\beta n \hbar \omega_0)/Z$, where the partition function 
$Z$ is $(1-\exp(\beta\hbar\omega_0))^{-11}$ and $\beta=1/k_B T$. 
Temperature enhances the absorption in the low energy regime 
and shifts the on-set 
of absorption to smaller energies. The temperature effect up to room 
temperature is very weak. At $T=300\rm{K}$ the probability to find a 
phonon on one of the eleven oxygens is less than 6\%. The analytical 
calculations for temperature corrections become increasingly tedious 
and Fig.\ref{fig:e2log} shows a first temperature correction obtained 
numerically.
%
\par

The clearest evidence for self-trapped excitons is normally
\cite{Song} observation of weak sub-gap features in optical
absorption $\alpha(\omega)\propto \epsilon_2(\omega)$ and
a large Stokes shift in the luminescence spectrum.  
We are not aware of luminescence measurements.  Transient
absorption measurements \cite{Federici} have been interpreted
as indicating that interband excitations relax rapidly to a
state shifted down by 0.7 eV.  Transmission measurements
through thin single crystals were reported by Kozlov {\it et al.}
\cite{Kozlov}.  These have been interpreted as showing an
indirect band gap at 0.5 eV with anomalous temperature
dependence.  We believe that the measured absorption can
be re-interpreted as an Urbach tail arising from the
self-trapped exciton, with binding energy $\epsilon_{\rm exc}$
enhanced beyond our values by the on-site Coulomb $U$.

Clear experimental support for the existence of self-trapped excitons 
in BaBiO$_3$
comes from harmonics of the Peierls breathing mode phonon
seen both in Raman experiments \cite{Tajima} and in
the infrared \cite{Kozlov}.  These spectra are
dominated by a 
strong peak at 570cm$^{-1}$, assigned to the breathing mode $Q=0$ $A_{1g}$
phonon.  Unlike in ordinary semi-conductors, a series of
higher harmonics is observed.  The simplest explanation 
for the Raman process is that the
incoming photon creates a virtual self-trapped exciton, except that 
because of the Franck-Condon principle, the lattice coordinates stay
at their ground state rather than their excited state values, which
means that a cloud of virtual local vibrations is also created.
When this virtual excitation re-radiates the scattered photon, 
not all the virtual vibrational quanta need to disappear.  The
test of this mechanism is that this process is resonant
when the laser photon energy coincides with the Peierls gap.
This resonance was seen by Tajima {\it et al.} \cite{Tajima}.
Similar behavior occurs in the Raman spectrum of LaMnO$_3$
\cite{Perebeinos} where a somewhat different species of self-trapped
exciton is predicted to occur \cite{Pereb2}.  In the infrared
spectrum of BaBiO$_3$, the
selection rule against Raman phonons appearing in infrared must
be violated by local loss of inversion symmetry due to static 
or dynamic lattice fluctuations.  Multiples of the breathing
phonon are as easily excited as a single phonon because of
the vibrational overlap factors.

\acknowledgements
We thank V. Perebeinos, V. Kostur, and R. Bhargava for help and encouragement.
This work was supported by NSF grant no. DMR-0089492.


\end{document}